\documentclass[final]{aipproc} 
\layoutstyle{6x9}

\begin{document}
\rightline{\vbox{\halign{&#\hfil\cr
			 &ANL-HEP-CP-02-097\cr}}}

\title{Determining Spin-Flavor Dependent Distributions}
\footnote{Talk given at SPIN 2002, Brookhaven National Laboratory, September 
9-13, 2002. E-mail address: gpr@hep.anl.gov} 
\footnote{Work supported by the U.S. Department of Energy, Division of
High Energy Physics, Contract W-31-109-ENG-38.} 

\author{Gordon P. Ramsey}{address={Loyola University Chicago and 
Argonne National Laboratory}}

\begin{abstract}
Many of the present and planned polarization experiments are focusing on 
determination of the polarized glue. There is a comparable set of spin 
experiments which can help to extract information on the separate 
flavor-dependent polarized distributions. This talk will discuss possible
sets of experiments, some of which are planned at BNL, CERN, DESY and JHF, 
which can be used to determine these distributions. Comments will include 
the estimated degree to which these distributions can be accurately found.
\end{abstract}

\maketitle

\section{Introduction}

During the past 20 years, considerable progress has been made in understanding
the nature of polarized distributions within nucleons. Various theoretical
models, coupled with data from polarized deep-inelastic scattering (PDIS) have
allowed extraction of polarized quark distributions. The net result is that
valence disctibutions and the up and down sea flavors are relatively well
determined, but the polarized strange sea, gluons and their corresponding
orbital angular momenta are unknown. The most recent efforts have generated
theoretical calculations and experiments at RHIC, HERA, and CERN, designed to 
determine $\Delta G$. Many of these experiments are in progress. \\

We now have the theoretical and experimental techniques to pursue more detail
into the flavor dependence of quark spin. Future efforts should include 
calculations and design of experiments to determine the spin contributions of 
all quark (and antiquark) flavors. This paper will discuss existing theoretical 
models, suggest a possible set of experiments and comment on the feasiblity of 
determining these distributions.

\section{Theoretical models}
	
Models for the spin contributions of the valence quarks are based mostly upon
modifications to the constituent quark model (CQM). \cite{CK,Isgur} These have 
the basic form:
\begin{eqnarray}
\Delta u_v(x)&=&M(x)[u_v(x)-2d_v(x)/3] \nonumber \\
\Delta d_v(x)&=&M(x)(-d_v(x)/3) \nonumber
\end{eqnarray}
where $M(x)$ is a modification factor to the CQM. In the Carlitz-Kaur model,
$M(x)$ is a "dilution" factor due to creation of gluons and the sea from 
valence quarks at small-$x$. In the Relativistic Constituent Quark Model
(Isgur), it represents a possible range of hyperfine interactions of the 
valence quarks with the other constituents. A statistical model, based upon 
the Pauli exclusion principle, \cite{B-S} generates a valence distribution in 
a similar form,
\begin{eqnarray}
\Delta u_v&=&u_v-d_v \nonumber \\
\Delta d_v&=&-d_v/3. \nonumber
\end{eqnarray}
These three models predict valence distributions that are qualitatively 
similar, but give a range of possible extrema for $\Delta u_v$ and $\Delta d_v$
in the valence region, which can be tested with suitable polarization 
measurements. 

The chiral quark model ($\chi$QM) \cite{Song} predicts integrals of the valence
distribution over $x$, with free parameters that can be fit with data. For 
appropriate ranges of these parameters, this model is consistent with the 
integral predictions of others. Lattice calculations of the moments of up and 
down valence distributions are consistent with the $\chi$QM for $\Delta u_v$,
but considerably less negative than the $\chi$QM prediction for $\Delta d_v$.
The exist a number of NLO fits of quark distributions to data, with assumed 
parametrizations of $\Delta u_v$ and $\Delta d_v$. \cite{B-T,LSS} These make 
certain assumptions about the symmetry of the polarized sea and could change
with more experimental information. All of these valence models are consistent
with the Bjorken Sum Rule and are similar in form. However, the differences 
are large enough to be distinguished by experimental measurements.

There is considerably more variance in the models for sea quarks, depending
upon the assumptions made about how the polarized sea is generated.
Sea models can be split into two categories: those based entirely on 
theoretical assumptions and the models that are a phenomenological combination
of theory and experimental data. We will consider models providing a completely
broken SU(3) polarized sea, where each flavor of quark/antiquark is determined 
separately. There is considerable theoretical evidence for broken SU(3).
\cite{WW,M-Y,KM} Lattice calculations of the moments of the up and down sea
have also indicated that this asymmetry could exist.

The statistical model mentioned above \cite{B-S} combines the Pauli exclusion 
principle, $F_2$ data and axial-vector couplings, $F$ and $D$ to represent the
polarized up quarks in terms of the unpolarized antiquarks. All other flavors 
are assumed to be unpolarized. This places a tight restriction on the size of 
the polarized sea. A light-cone model of meson-baryon fluctuations puts the
intrinsic $q\bar{q}$ pairs with the valence quarks in an energetically favored
state. \cite{BM} In this model, coupling to virtual $K^+ \Lambda$ hyperons is 
the source of intrinsic $s\bar{s}$ pairs. Thus, the antiquarks are unpolarized
and the light flavored quarks are polarized opposite to that of the proton.

In the chiral quark model, \cite{Song} the polarized sea determined by chiral 
fluctuations of the valence quarks, creating Goldstone bosons, which result in
the prediction that $\Delta \bar{q}=0$ all flavors. As with the valence quarks, 
ranges for the integrals of the polarized sea quarks are predicted. The meson 
cloud model is similar, with pseudo-scalar mesons replacing the Goldstone 
bosons. \cite{KM} In contrast to the $\chi$QM, the result is that 
$\Delta \bar d=\Delta u$ and the remaining quarks are unpolarized. In a chiral 
quark-soliton model, \cite{WW} quark fields interact with massless pions,
yielding an asymmetry for the polarized up and down quarks, related to the
unpolarized up and down antiquarks. This model has been phenomenologically
tested in polarized semi-inclusive processes. \cite{M-Y}

Most predictions of heavy quark contributions to proton spin indicate that
they are likely small. The $\chi$QM prediction gives $\Delta c\approx -0.003$
and $\Delta \bar{c}=0$. Similarly, an analysis using the operator product 
expansion and the axial anomaly predicts that $\Delta c=-0.0024\pm 0.0035$,
consistent with the $\chi$QM.\cite{man} Instanton models tend to predict 
somewhat larger contributions from the heavier quarks. \cite{bs} These range 
from $\Delta c=-0.012\pm 0.002\to -0.020\pm 0.005$. Thus, $\Delta c$ is at 
most a very small fraction of the total polarized sea and will likely prove 
quite difficult to measure.

Table 1 contains key results from some of the models described above for
comparison. The most significant differences are in the predictions for the 
polarization of the antiquarks. This distinction can also be carried over to 
the theoretically motivated phenomenological models.

\begin{table}
\begin{tabular}{c|cccccccc}
\hline
Model & $\Delta u$ & $\Delta \bar{u}$ & $\Delta d$ & $\Delta \bar{d}$ &
$\Delta s$ & $\Delta \bar{s}$ & $\Delta c$ & $\Delta \bar{c}$ \cr
\hline\hline
Statistical & $\bar{u}-\bar{d}$ & $\bar{u}-\bar{d}$ & 0 & 0 & 0 & 0 & 0 & 0 \cr
\hline
L-Cone & -- & -- & $< 0$ & 0 & $< 0$ & 0 & 0 & 0 \cr
\hline
$\chi$QM & 0.83 & 0 & -0.39 & 0 & -0.07 & 0 & -0.003 & 0 \cr
\hline
M-cloud & $\Delta u$ & 0 & 0 & $\Delta u$ & -- & -- & 0 & 0 \cr
\hline
CQSM & -- & $\Delta \bar{d}-Cx^\alpha(\bar{d}-\bar{u})$ & -- & see $\Delta 
\bar{u}$ & 0 & 0 & 0 & 0 \cr
\hline
\end{tabular}
\caption{Sea flavor contributions by type of model}
\label{tab:1}
\end{table}

Phenomenological models range from those grounded in theoretical constraints
and use data to fit parameters to the ones which are primarily parametrizations
determined by fits to data. Models in which the polarized sea is 
created by gluons, that pass polarization "information" to the quarks by the 
splitting process, are in the former category. \cite{B-T,G-G-R} In the GGR 
model, \cite{G-G-R} the flavor asymmetry of the polarized sea is caused by the 
asymmetry in the unpolarized distributions. Specific forms for the 
parametrization of the separate distributions come from axial-vector 
constraints and data. 

Most direct data fits \cite{B-T,LSS} assume minimal SU(3) breaking of the 
polarized sea. Similarly, LO/NLO moment fits to data \cite{B-B} result in only
a small amount of SU(3) breaking, but a stronger asymmetry of the sea is
possible within the cited error analysis. These theoretical and 
phenomenological models provide a sufficient variance for experiments to be 
able to distinguish between their fundamental assumptions.

\section{Experiments}

Polarized valence distributions can be fine-tuned by measuring asymmetries in 
pion production. By taking differences of these asymmetries for $\pi^+$ and
$\pi^-$ production, the valence contributions can be extracted. This results 
in: 
\begin{equation}
\Delta A^{\pi}\equiv A_p^{\pi^+}-A_p^{\pi^-}={{4\Delta u_v-\Delta d_v}\over
{4u_v-d_v}}.
\end{equation}
In the valence models previously discussed, these asymmetries differ by 
$0.2$ for $x<0.5$ and by $0.1$ for $0.5\le x\le 0.9$. Similarly, differences 
in $\pi^0$ production for $p$ and $\bar p$ yield large asymmetries for 
$0.1\le p_T/\sqrt(s)\le 0.3$, but high energy $\bar p$ beams with sufficient 
luminosity for good statistics are difficult to achieve. \\

Present measurements of $\Delta (q+\bar q)/(q+\bar q)$ for the light quark 
flavors at HERA are providing a good start at finding the contributions of
these flavors to the spin of the proton. \cite{HERA} We would like to determine
the individual spin contributions of each quark and antiquark flavor. For this, 
a combination of polarization experiments will be necessary.
Charged current interactions are a useful tool in investigating kinematic
dependences of both the polarized valence and sea quark distributions. The 
single spin asymmetries in parity-violating $W$ production ($A_L^{W^\pm}$)
can yield valuable information about the polarization of light quark flavors.
\cite{bsf}
\begin{equation}
A_L^{W^+}(y)={{\Delta u(x_a)\bar d(x_b)-\Delta \bar d(x_a)u(x_b)}\over
{u(x_a)\bar d(x_b)+\bar d(x_a)u(x_b)}}
\end{equation}

\begin{equation}
A_L^{W^-}(y)={{\Delta d(x_a)\bar u(x_b)-\Delta \bar u(x_a)d(x_b)}\over
{d(x_a)\bar u(x_b)+\bar u(x_a)d(x_b)}}
\end{equation}
For example, at $y=0$, $x\approx M_W/\sqrt{s}$ and the asymmetry measures 
combinations of $u$ and $\bar d$ or $d$ and $\bar u$. 
For $y=-1$, $x$ is small and the second terms in each numerator and denominator
dominate, so we can separately probe $\bar u$ and $\bar d$.
At $y=+1$, $x$ is of moderate value, the first terms in each numerator and 
denominator dominate so that both $u$ and $d$ polarizations can be measured. 
This would provide a more complete picture of the light quark polarizations.
However, a limited kinematic range will be probed at RHIC. Therefore, this
should be combined with other experiments to probe the sea polarization.

Combinations of polarized sea flavors can be investigated in a number of
different experiments. For $W^{\pm}$ production in the HERA kinematic range,
$g_5/F_1$ is extracted from the measured asymmetry. This yields the following
combinations:
$g_5^{W^-}=\Delta u+\Delta \bar d+\Delta \bar s+\Delta c$ and
$g_5^{W^+}=\Delta \bar u+\Delta d+\Delta s+\Delta \bar c$.
However, the uncertainties in the hadronic energy scale of the calorimeter are
of comparable size to the asymmetries. Measurements may be difficult at RHIC 
as well, since these asymmetries are generally small at its kinematic range.
Measurement of $g_1$ in polarized $e^{\pm}p\to \nu (\bar \nu)X$ scattering at 
HERA could yield a similar combination of polarized flavors.

Parity-violating $\nu$ scattering ($p$ and $n$) measurement of $g_3$ at the
proposed Japan Hadron Facility would give:
$\frac{1}{2}[g_3^{\nu(p+n)}-g_3^{\bar \nu(p+n)}]\sim \Delta s+\Delta \bar s-
\Delta c-\Delta \bar c$.
However, since $g_3$ comes from $W_{\mu\nu}^{\perp}$, which is small, this may
be difficult to distingish. \\

Parity conserving double spin asymmetries in $Z$ production ($A_{LL}^{Z^0}$)
provide a valuable tool in investigating the polarization of the 
antiquarks. This asymmetry is given by:
\begin{equation}
A_{LL}^{Z^0}(y)\sim \Sigma_i {{\Delta q_i(x_a)\Delta \bar q_i(x_b)+\Delta 
\bar q_i(x_a)\Delta q_i(x_b)}\over {q_i(x_a)\bar q_i(x_b)+
\bar q_i(x_a)q_i(x_b)}} \nonumber
\end{equation}
Predicted asymmetries of $\sim 0.10$ for $\sqrt(s)=500$ GeV could be 
distiguishable from zero with 400-500 events at RHIC. This would provide an 
excellent test of the light-cone, $\chi$QM and instanton models that predict 
zero polarization for antiquarks.

Polarized Drell-Yan experiments at both RHIC (at 50-100 GeV) or the proposed 
Japan Hadron Facility (JHF) at 50 GeV provide promising ways to extract more
precise information about the polarization of the sea. At these energies, the
cross sections are larger and the asymmetries are moderately sized. This makes
the competing predictions easy to distinguish. The cross sections decrease
rapidly with energy, so experiments at larger $\sqrt{s}$ are not good
candidates for this set of measurements. The RHIC luminosity is low at 50 GeV, 
(the injection energy) but probably suitable at 100 GeV. Polarized beams at the
JHF are quite appropriate for lepton pair production experiments and would be
excellent for determining the relative size of the polarized sea. \cite{JHF}
These measurements in principle could distinguish the flavor dependence of the
polarized sea. Combined with the experiments described above, they would give a 
complete picture of the sea polarization.

\section{Conclusion}

There has been considerable progress in narrowing the polarizations of the 
lighter quark flavors, $\Delta q_v$, $\Delta u_{tot}$ and $\Delta d_{tot}$.
There exist many theoretical predictions for polarizations of the valence, sea
quark and antiquark flavors. The experiments described here include most 
possibilities for determining the spin contributions of four quark and
antiquark flavors. Many of the suggested measurements are feasible and should 
be done, since a combination of experiments would give the best range of
information about quark spin. This opportunity opens up numerous possibilities
for polarization experiments at RHIC, HERA, COMPASS and the JHF.


\begin{thebibliography}{0}
\bibitem {CK} Carlitz and Kaur, Phys. Rev. Lett. \underline{38}, 673 (1977).
\bibitem {Isgur} Isgur, N. Phys. Rev. \underline{D59}, 034013 (1999) and 
hep-ph/9809255.
\bibitem {B-S} Bourrely and J. Soffer, Nucl. Phys. \underline{B445}, 341 (1995)
and Eur. Phys. J. \underline{C23}, 487 (2002).
\bibitem {Song} X. Song, Phys. Rev. \underline{D57}, 4114 (1998).
\bibitem {B-T} Bartelski and Tatur, Phys. Rev. \underline{D65}, 034002 (2002) 
and hep-ph/0107202.
\bibitem {LSS} E. Leader, Siderov, Stamenmov, Eur. Phys. J. \underline{C23},
479 (2002) and hep-ph/0111267.
\bibitem {WW} Wakamatsu and Watabe, Phys. Rev. \underline{D62}, 017506 (2000).
\bibitem {M-Y} T. Morii and Yamanishi, Phys. Rev. \underline{D61}, 057501
(2000), erratum: Phys. Rev. \underline{D62}, 059901 (2000) and D. de Florian
and R. Sassot, Phys. Rev. \underline{D62}, 094025 (2000).
\bibitem {KM} S. Kumano and Miyama, Phys. Rev. \underline{D65}, 034012 (2002).
\bibitem {BM} S. Brodsky and B-Q. Ma, Phys. Lett. \underline{B381}, 317 (1996).
\bibitem {man} A. Manohar, Phys. Lett. \underline{B242}, 94 (1990).
\bibitem {bs} Blotz and Shuryak, Phys. Lett. \underline{B439}, 415 (1998).
\bibitem {G-G-R} L. Gordon, M. Goshtasbpour and G. Ramsey, Phys. Rev. 
\underline{D58}, 094017 (1998).
\bibitem {B-B} J. Bl\"umlein and B\"ottcher, Nucl. Phys. \underline{B636}, 225 
(2002).
\bibitem {HERA} See \url{http://www.desy.de}
\bibitem {bsf} Bourrely and Soffer, Phys. Rev. \underline{D51}, 2108 (1995).
\bibitem {JHF} J. C. Peng, {\it et. al.}, hep-ph/0007341 and S. Kumano, 
hep-ph/0207151. 
\end{thebibliography}
\end{document}